# Enhancing Cyber-Resilience in Cyber-Physical Systems of Systems: A Methodical Approach

Elisabeth Vogel[1,2], Peter Langendörfer[1,2]

[1]Leibniz Institute for High Performance Microelectronics (IHP), Frankfurt (Oder), 15236, Germany
[2]Chair of Wireless Systems, BTU Cottbus-Senftenberg, Cottbus, 03046, Germany

*Correspondence to*: Elisabeth Vogel (Elisabeth.Vogel@b-tu.de)

**Abstract.** Cyber-physical systems of systems (CPSoS) are becoming increasingly prevalent across sectors such as Industry 4.0 and smart homes, where they play a critical role in enabling intelligent, interconnected functionality. Addressing the challenges and resilience requirements of these complex environments, we propose a modified Cyber-Resilience Life-Cycle as a practical framework for sustainable risk mitigation. Our approach enhances the adaptability of CPSoS and supports resilience against evolving system complexities and potential disruptions. We conclude by outlining application scenarios for the modified life-cycle and highlighting its relevance in fostering cyber-resilience in operational systems.

## 1 Introduction

Cyber-Physical Systems of Systems (CPSoS) permeate almost all areas of human life. We define a CPSoS as follows: CPSoS are highly complex systems of systems consisting of the interaction of physical components (e.g., sensors, actuators) with information technology and communication technology.

In the public sector, CPSoS now play a central role, having already revolutionised many areas. These include, for example, Industry 4.0, where the networking of machines and production facilities has greatly increased efficiency and automation. Fundamental progress has also been made in the transport sector, with the spread of autonomous driving. In the energy industry, smart grids are contributing to more sustainable and efficient energy use. Other examples of applications include the aerospace industry, where CPSoS promotes innovation and safety, and telecommunications, where it forms the basis for the next generation of communication technologies (Carías et al., 2020; Hopkins et al., 2020; Linkov and Kott, 2019; Yu Wang et al., 2023).

CPSoS are also becoming increasingly popular in the private sector. Everyday tasks can be automated with the help of smart home technologies. Smart wearables are able to capture and analyse personal health and fitness data in real time. Intelligent household appliances are constantly being improved in terms of both functionality and efficiency, enabling them to support the user. At the same time, personal assistants and intelligent speakers allow easier and more efficient communication with the digital world. This is achieved by integrating voice control and AI-based functions into everyday life. These developments impressively show that CPSoS are not only becoming increasingly popular in the industrial and technological landscape, where it has already achieved major changes. CPSoS are also increasingly permeating and improving our private lives.

Depending on the application, CPSoS are subject to many different requirements and challenges. These requirements and challenges differ considerably from sector to sector. The most important requirements include reliability, availability, security against (cyber-) attacks and disturbances, interoperability between different systems, real-time capability, flexibility in adapting to different deployment scenarios, and scalability. Depending on the application, these requirements vary considerably. The consequences of not meeting certain requirements in a (highly) critical environment, such as a nuclear power plant, a dam or a hospital, are significantly more severe than in a private environment. The risks and effects in a private environment are significantly more manageable.

The ever-growing and evolving list of requirements and challenges can be traced back to the ubiquitous presence and importance of CPSoS. With continuous technological advancements and the increasing integration of such systems into new areas, not only does increase their complexity, but so does the need to meet ever more demanding and industry-specific requirements. This continuous development presents engineers and developers with the task of developing solutions that meet the unique requirements of each area without compromising the safety and efficiency of the systems.

The concept of cyber resilience offers a holistic solution to address these growing challenges. This paper presents the fundamental conditions that CPSoS must meet to comply with the concept of cyber resilience. From these conditions, a model-based approach can be derived that can be used to cyclically manage the mitigation of disturbances and risk factors of CPSoS. In Figure 1 [a], we present the cyber resilience lifecycle described in (Vogel et al., 2021). The Cyber-Resilience Life-Cycle shown in Figure 1 [a] is adapted from a similar cycle developed by (Rosati et al., 2015), which was originally designed to quantify the resilience of coastal systems and areas following flood events.

In the course of our research, we identified several weaknesses in our original adaptation of the life cycle (Figure 1 [a]). Therefore, we revised our Cyber-Resilience Life-Cycle based on the model by (Vogel et al., 2021), as shown in Figure 1 [b]. These revisions include: 1) the rearrangement of the adaptation and recovery phases, 2) the redefinition of resistance, and 3) the explicit definition of a disturbance in the context of a CPSoS.

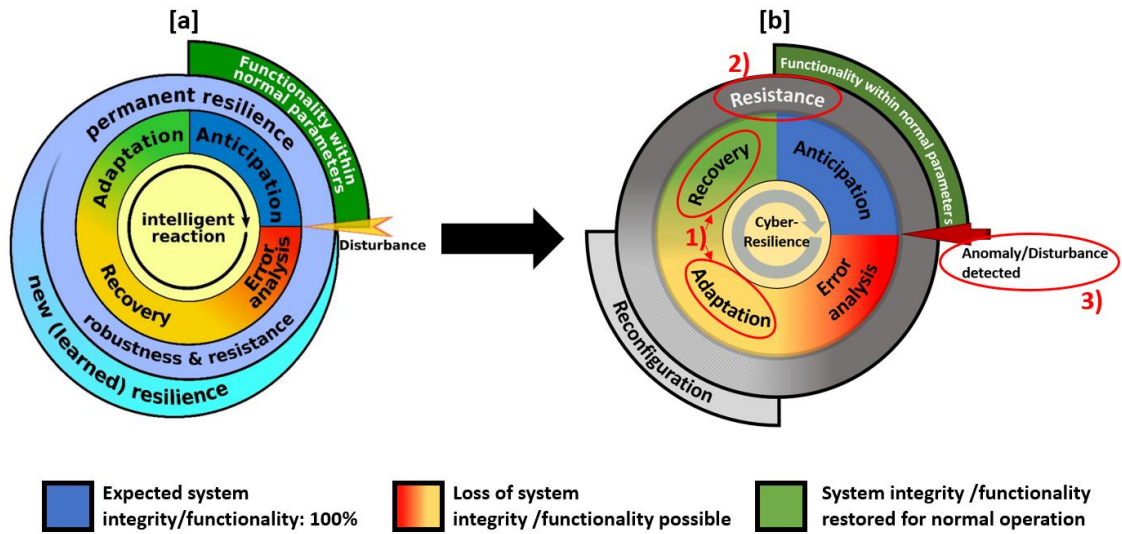

**Figure 1:** Representation of [a] the original Cyber-Resilience Life-Cycle (Vogel et al., 2021) and [b] the modified Cyber-Resilience Life-Cycle. The modified version reflects key conceptual changes, including (1) the rearrangement of the adaptation and recovery phases, (2) a redefinition of the resistance, and (3) an updated understanding of disturbances and abnormal system behaviour in the context of CPSoS.

The fundamental principle of resilience is not to avoid disturbances, but to be able to respond to them appropriately and to learn from them. The aforementioned modifications to our Cyber-Resilience Life-Cycle reflect this principle and are intended to improve the system's ability to handle disturbances more efficiently. To introduce these changes, we first present our original Cyber-Resilience Life-Cycle (Figure 1[a]) in Section 2. This section includes definitions of key terms such as CPSoS and the main components of the Cyber-Resilience Life-Cycle. In Section 3, we present our revised version of the Cyber-Resilience Life-Cycle (Figure 1 [b]). The specific changes are explained and justified in detail, along with a discussion of how they improve the overall concept. Finally, in Section 4, we summarise our results and provide an outlook on possible applications of the Cyber-Resilience Life-Cycle. We also outline potential next steps toward a resilience-oriented system design.

## 2 Background

CPSoS are sometimes highly complex systems that combine a large number of functions and processes to connect physical and digital components. Their high level of complexity makes these systems vulnerable to a range of risk factors that can affect both the functionality and the integrity of the affected system. The risk factors that CPSoS face can be roughly categorised into two main categories as follows:

1) Internal risk factors

    Internal risk factors are usually software errors or hardware failures that can occur within the system. These errors can be caused by inadequate/erroneous programming, faulty components or unforeseen interactions between different system elements.

2) External risk factors

    On the other hand, there are external risk factors that are physically outside the system. These can significantly impair functionality. External risk factors include, among other things, human error or failure, such as operating errors or misunderstandings. This category also includes cyber-attacks (Michael J. Weisman et al., 2023; Dyka et al., 2020; Dyka et al., 2019) that aim to compromise the security and/or integrity of CPSoS data and processes.

As the complexity of CPSoSs increases, not only do the risk factors themselves become more extensive, but so do their interactions. The more different system components encounter different risk factors, the more clearly dependencies and interactions emerge. This complication of the risk landscape must be taken into account.

The directly observable effects of risk factors that occur often manifest themselves as disturbances in system operation. These disturbances can be divided into two main categories:

1) Visible disturbances

    Visible disturbances are those that are easily detected and lead to an immediate loss of functionality. Examples include (partial) hardware failures, which become apparent through the malfunctioning of devices or components. Such failures are often relatively easy to diagnose and repair, as they are usually directly connected to a specific physical element of the system.

2) Invisible disturbances

    Invisible disturbances are much more subtle and can seriously compromise the integrity of a CPSoS without any immediate or obvious signs. A notable example of this are cyber-attacks, which aim to damage a system or steal sensitive data. This type of threat often remains hidden until it has already caused significant damage or compromised critical information.

To create a cyber-resilient CPSoS, it is crucial to empower the CPSoS to quickly and reliably detect both visible and invisible disturbances, understand their causes and react appropriately. Such resilience enables the system not only to overcome the challenges and threats posed by these disturbances, but also to adapt to new conditions and quickly return to normal operation. In line with these requirements, we have defined resilience as follows (Vogel et al., 2021):

*"A CPS(oS) is resilient if it has the ability to react to specified and unspecified disturbances in a way that preserves its function and reacts quickly. This reaction includes the early detection, minimization, prediction or even avoidance of disturbances. In addition, it needs to have the capability to anticipate future challenges and to prepare itself for those."*

In line with this definition of the concept of cyber resilience, the core idea is not to prevent disturbances. The reason for this is that today's CPSoS are (1) extremely complex and (2) exposed to an equally complex and dynamic range of disturbances. It is unrealistic to assume that this system complexity and the complexity of potential disturbances can be perfectly aligned in a way that would allow all possible disruptions to be completely prevented. The goal is therefore to ensure that disturbances are handled in such a way that the system's function and integrity are maintained to the greatest extent possible. Due to their enormous complexity, disturbances of highly complex CPSoS cannot be prevented, but their negative effects on the CPSoS should be minimised by appropriate preventive measures.

These response strategies should aim to minimise the negative effects of disturbances and maintain system performance, ideally during and in any case after a disturbance. This includes proactive preventive measures to identify and mitigate potential risks. For example, redundant systems, regular maintenance and updates, staff training and the implementation of security protocols can help to reduce the probability and severity of possible disturbances.

The definition of cyber resilience describes a set of characteristics or capabilities (Vogel et al., 2021) that a system must possess to be considered cyber-resilient. These characteristics, referred to as key actions (Rosati et al., 2015; Vogel et al., 2021), are essential for responding effectively to disturbances and maintaining the functionality and integrity of the system. Table 1 provides a detailed description of these key actions.

**Table 1: This table provides an enumeration and description of the key actions essential for the effective management and resilience of CPSoS.**

| Key action | Description |
|---|---|
| Anticipation | The ability to predict future events by taking into account already known information. |
| Error analysis | The ability to identify and interpret disturbances. |
| Recovery | The ability to restore a system's functionality within acceptable parameters after an incident. |
| Adaptation | The ability of a system to perform a self-modification in order to transition to a new state so that the system can respond more effectively to the same or similar disturbance(s) in the future. This results in less loss of function and/or shorter recovery times. |
| Permanent resilience | Permanently active countermeasures for known and/or expected disturbances. Disturbances are expected if they are foreseeable for the system under consideration. |
| New (learned) Resilience | Countermeasures that are not permanently active but can be deployed or activated as the situation requires. |

The key actions listed in Table 1 are not static but time-dependent, and they form an action cycle that gives the cyber-physical system of systems (CPSoS) clear instructions on how to respond to the effects of a disturbance. For this action cycle to function successfully, it is essential that the CPSoS is able to detect a disturbance in good time. Figure 1 illustrates this action cycle, which is known as the 'cyber resilience life cycle'. This cycle shows how a CPSoS reacts systematically to disturbances and adapts to maintain the integrity and functionality of the system and return to normal operation as quickly as possible.

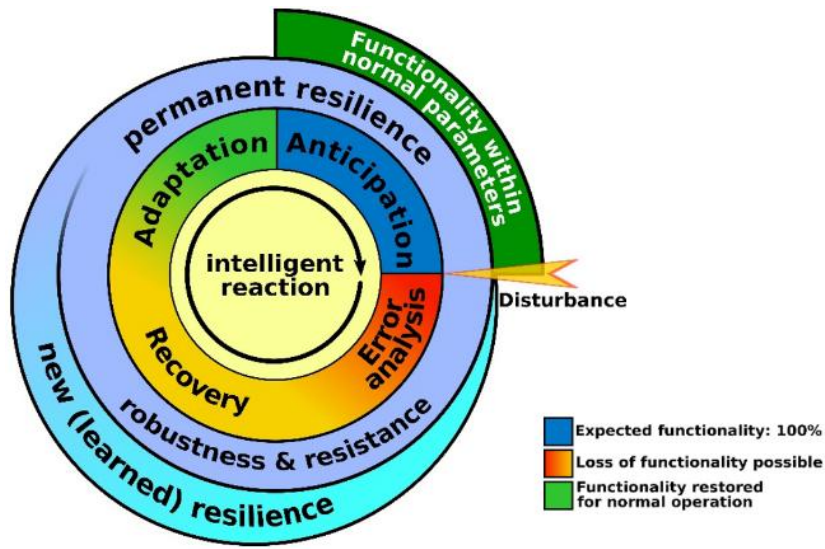

**Figure 2: The Cyber-Resilience Life-Cycle represents the series of key actions and system states that enable a CPSoS to anticipate, respond to, and recover from disturbances while continuously adapting to enhance future resilience. (Vogel et al., 2021)**

The Cyber-Resilience Life-Cycle depicted in Figure 2 visualises the key actions described in Table 1 within an interdependent cycle. In a broader sense, these key actions can also be described as system states, since each key action causes a specific behaviour of the system:

By default, CPSoS is in the system state "Anticipation", which represents the state of near-optimal functionality and integrity of the system. In this state, the system is able to predict potential disturbances through various measures, as well as introspection and analysis. This ability enables CPSoS to take appropriate precautions to maintain functionality and integrity in the face of possible future disturbances.

When a disturbance occurs that is either unknown to the system or for which no suitable course of action has yet been defined, the CPSoS enters the system state "Error analysis". Entry into this state requires the early detection of the disturbance. This is an essential prerequisite for further analysis. In the error analysis state, the disturbance is analysed, its negative effects on the system are evaluated, and the available countermeasures are checked. The aim is to understand the disturbance and its potential effects on the system as precisely as possible. Only then is it possible to select suitable strategies or strategy combinations to restore system integrity and functionality efficiently and quickly.

In the "recovery" system state, any lost system functionality and integrity is restored as far as possible. Various factors are crucial to the success of the restoration of lost system functionality and integrity. In this context, the extent of the damage caused by the disturbance is of particular importance. For example, if substantial physical components have to be replaced because redundancy is either not available or exhausted, a complete restoration is not possible.

When transitioning to the "adaptation" system state, the modified countermeasures are integrated into the existing countermeasures and updated. This is shown schematically in Figure 2 as a combination of the countermeasures already in use (robustness and resilience) with the new modifications (new (learned) resiliency) with regard to the disturbance that has just occurred. The combination of these results in the new adaptation of the countermeasures with regard to the disturbance that has occurred (permanent resilience).

In the future, CPSoS will be able to anticipate this disturbance and minimise any functional losses that may occur. The more frequently this disturbance occurs, the more efficient this adaptation will be.

## 3 Modified Cyber-Resilience Life-Cycle

Based on our extensive research into cyber resilience and its relevance to CPSoS, we propose a fundamental revision of the Cyber-Resilience Life-Cycle described in Section 2.

To begin this refinement, attention must be paid to the key actions described in Table 1. The inner cycle, consisting of anticipation, failure analysis, recovery and adaptation, remains untouched with respect to the definitions and associated system states.

A crucial aspect, however, is the re-evaluation of the chronological order of the key actions in the inner cycle. Following the approach in (Vogel et al., 2021), we assume that a CPSoS initially exists in a state that allows for the prediction of potential future failures. Upon failure detection, the CPSoS transitions to failure analysis, preserving the essence of the previous cyber resilience lifecycle (see Figure 2). However, the advancement in the revised model is that the CPSoS initiates decisive changes to active countermeasures before it begins to restore functionality and/or integrity. This strategic shift is essential to avoid potential damage at both the hardware and software levels. A system must first make the necessary modifications in response to the disturbance it has encountered and can only then begin to restore its system functionality or integrity.

If this change between adaptation and recovery is not taken into account, there is a risk that the system will make modifications based on an incorrect database in the form of an incorrect system environment. Furthermore, there is a risk that the effects of potential disturbances will no longer be recognised as such, but could be interpreted as the system standard. Such effects could then be propagated over many cycles of the cyber-resilience life cycle because they can no longer be recognised. The result could be a comprehensive cascade of errors that can no longer be reproduced and can cause extensive damage, especially in large clusters of systems.

A second change that we consider essential is the consideration of the implementation of new modifications and their integration into the countermeasures. Figure 2 shows that new countermeasures, corresponding to the disturbance under consideration, should be integrated into the pool of already active countermeasures. This combination then results in a number of countermeasures (permanent resilience) that are relevant for the disturbance under consideration. This implementation is possible for simple or small systems, but it does not reflect the complexity of CPSoS, which is used as standard at the time of publication of this paper. Therefore, we advocate combining the key actions 'permanent resilience' and 'new (learned) resilience' and refer to this connection as 'resistance' in the future. The visual representation of these modifications is shown in Figure 3 and indicates a significant development in the life cycle of cyber resilience.

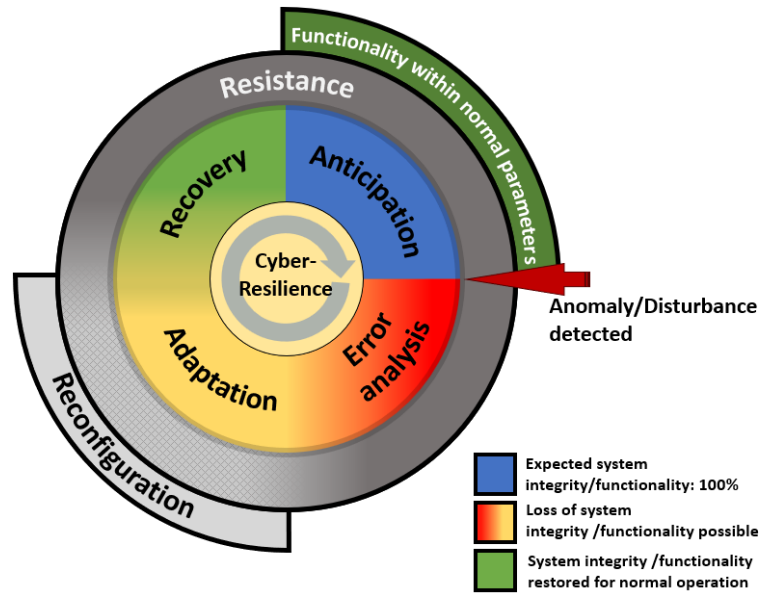

**Figure 3: The illustration shows the Cyber-Resilience Life-Cycle with the necessary adjustments. The model shown here includes proactive adjustments to countermeasures before recovery from an incident. This approach is essential to minimise damage and avoid cascading failures.**

As shown in Figure 3, the resistance now completely encases the inner cycle. When the system is in the adaptation state, adjustments are made to all relevant countermeasures or their usefulness is reviewed based on the current data. There are often interdependencies between different countermeasures, the effects of which can also have negative consequences for the system. These dependencies must be found, evaluated and taken into account accordingly.

The efforts to minimise damage are therefore very comprehensive. For example: In scenarios with redundant system components, a practical solution is to switch to unaffected components. However, if significant damage occurs that exceeds the recovery capabilities of CPSoS, human intervention is essential. It is of utmost importance, especially in the context of a cyber-resilient system, to minimise the dependency on human intervention. Consequently, the strategic integration of appropriate redundancies and the well-thought-out coupling or separation of system components in the initial stages of the design phase proves to be a critical aspect.

Finally, we propose a differentiated definition of disturbances. We argue that it is useful to distinguish between abnormal behaviour and specific disturbances. If a system is able to detect abnormal behaviour, appropriate countermeasures can be initiated before a disturbance occurs. In this sense, abnormal behaviour can be considered a precursor to a disturbance. At the same time, it may also be observed during a disturbance, indicating that both states are closely interrelated. Abnormal behaviour is therefore a key factor in enabling fast and efficient responses, particularly in the case of known or anticipated disturbances.

## 4 Conclusion

In our view, the modification of the cyber resilience life cycle described here is a necessary further development of the variants from (Vogel et al., 2021; Rosati et al., 2015).

The cycle of key actions described in (Rosati et al., 2015) deals with the transition from risk management to disaster resilience and refers to the management of water resource infrastructure in the United States. Factors such as changes in climate patterns, increasing environmental concerns, higher population density in coastal areas, related infrastructure, limited budgets and aging infrastructure were considered as risk factors. (Rosati et al., 2015) points out that risk management, as the predominant assessment strategy for engineered systems, aims to provide solutions for known hazards in terms of continued normal operation. When uncertainties arise in technical systems, risk management assumes that these can be quantified and mitigated (Park et al., 2013). However, the more complex a system becomes, the more influencing factors such as feedback loops, interactions, (future) unknowns and variable spatial or temporal scales for hazards arise. Many of these factors are also unpredictable (Rosati et al., 2015).

For this reason, a solution for the transition from risk management to a resilience approach has been proposed as an alternative, since resilience pursues the handling and processing of the unknown and the unforeseeable as a guiding principle.

For the design of the cyber resilience life cycle (Vogel et al., 2021), we have adopted the approach from (Rosati et al., 2015) and initially adapted it for cyber-physical systems (see Figure 2). Similar to the transition from risk management to disaster resilience, we consider a shift from conventional fault tolerance to a resilience approach to be necessary.

The increasing complexity of cyber-physical systems is another factor that must be taken into account. Conventional fault tolerance techniques (Bharany et al., 2022) may no longer be sufficient. The risk factors described in (Rosati et al., 2015) are very diverse and come from a wide range of areas. A CPSoS is exposed to a similar range of risk factors.

With a growing understanding of resilience approaches in CPSoS, we have now modified our Cyber-Resilience Life-Cycle from (Vogel et al., 2021) according to Figure 3. The Cyber-Resilience Life-Cycle has been reorganised by changing the order of the recovery and adaptation phases.

From our point of view, it is essential in the context of CPSoS that a system modifies its countermeasures based on the error analysis. This must be done after the error analysis and before functionality or integrity can be restored. Doing so prevents the CPSoS from being in a weakened or even vulnerable state during recovery, which would make it more susceptible to disturbances and help avoid cascading effects. With the modification of the Cyber-Resilience Life-Cycle from (Vogel et al., 2021), the CPSoS can make all the necessary adjustments and implement improvements to restore functionality and integrity, without increasing the risk of a new failure.

The designed and modified theoretical model of the cyber-resilience life cycle can now be used as a basis for the implementation of a concrete model to determine the resilience of a CPSoS. With the help of such a model, a general metric for the resilience of a CPSoS can be created first, and a specific metric can then be created based on this.

**Author Contributions**

Conceptualization, EV and PL; methodology, EV and PL; validation, PL; formal analysis, EV and PL; writing—original draft preparation, EV; writing—review and editing, PL; visualization, EV; supervision, PL
All authors have read and agreed to the published version of the manuscript.

**Competing interests**

The authors declare that they have no conflict of interest.

**References**

Bharany, S., Badotra, S., Sharma, S., Rani, S., Alazab, M., Jhaveri, R. H., and Reddy Gadekallu, T.: Energy efficient fault tolerance techniques in green cloud computing: A systematic survey and taxonomy, Sustainable Energy Technologies and Assessments, 53, 102613, doi:10.1016/j.seta.2022.102613, 2022.

Carías, J. F., Borges, M. R. S., Labaka, L., Arrizabalaga, S., and Hernantes, J.: Systematic Approach to Cyber Resilience Operationalization in SMEs, IEEE Access, 8, 174200–174221, doi:10.1109/ACCESS.2020.3026063, 2020.

Dyka, Z., Vogel, E., Kabin, I., Aftowicz, M., Klann, D., and Langendorfer, P.: Resilience more than the Sum of Security and Dependability: Cognition is what makes the Difference, in: 2019 8th Mediterranean Conference on Embedded Computing (MECO): Including ECYPS '2019 proceedings-research monograph Budva, Montenegro, June 10th-14th, 2019, Stojanovic, R. (Ed.), 2019 8th Mediterranean Conference on Embedded Computing (MECO), Budva, Montenegro, 6/10/2019 - 6/14/2019, IEEE, Piscataway, NJ, 1–3, 2019.

Dyka, Z., Vogel, E., Kabin, I., Klann, D., Shamilyan, O., and Langendörfer, P.: No Resilience without Security, in: 2020 9th Mediterranean Conference on Embedded Computing (MECO), 1–5, 2020.

Hopkins, S., Kalaimannan, E., and John, C. S.: Foundations for Research in Cyber-Physical System Cyber Resilience using State Estimation, in: 2020 SoutheastCon, 1–2, 2020.

Linkov, I. and Kott, A.: Fundamental Concepts of Cyber Resilience: Introduction and Overview, in: Cyber Resilience of Systems and Networks, Kott, A., and Linkov, I. (Eds.), Springer International Publishing, Cham, 1–25, 2019.

Michael J. Weisman, Alexander Kott, Jason E. Ellis, Brian J. Murphy, Travis W. Parker, Sidney Smith, and Joachim Vandekerckhove: Quantitative Measurement of Cyber Resilience: Modeling and Experimentation, arXiv preprint arXiv:2303.16307, 2023.

Park, J., Seager, T. P., Rao, P. S. C., Convertino, M., and Linkov, I.: Integrating risk and resilience approaches to catastrophe management in engineering systems, Risk analysis an official publication of the Society for Risk Analysis, 33, 356–367, doi:10.1111/j.1539-6924.2012.01885.x, 2013.

Rosati, J. D., Touzinsky, K. F., and Lillycrop, W. J.: Quantifying coastal system resilience for the US Army Corps of Engineers, Environ Syst Decis, 35, 196–208, doi:10.1007/s10669-015-9548-3, 2015.

Vogel, E., Dyka, Z., Klann, D., and Langendörfer, P.: Resilience in the Cyberworld: Definitions, Features and Models, Future Internet, 13, doi:10.3390/fi13110293, 2021.

Yu Wang, Chao Deng, Yun Liu, and Zhongbao Wei: A cyber-resilient control approach for islanded microgrids under hybrid attacks, International Journal of Electrical Power & Energy Systems, 147, 108889, doi:10.1016/j.ijepes.2022.108889, 2023.